%% file: hypercube7.tex
\documentstyle[preprint,tighten,aps,amstex,graphicx,times,floats]{revtex}

\begin{document}

\thispagestyle{empty}
 
\input{declare}
\preprint{
\font\fortssbx=cmssbx10 scaled \magstep2
\hbox to \hsize{
%\hskip.5in \raise.1in\hbox{\fortssbx University of Wisconsin - Madison}
\hfill\vtop{\hbox{\bf MADPH-03-1344}
            \hbox{\today}                    } }
}

\title{IceCube-Plus: An Ultra-High Energy Neutrino Telescope}

\author{Francis Halzen$^1$ and Dan Hooper$^{1,2}$}

\address{
$^1$ Department of Physics, University of Wisconsin, 1150 University Avenue, Madison, WI  53703, USA \\
$^2$ Denys Wilkinson Laboratory, Astrophysics Department, OX1 3RH Oxford, England, UK}

\maketitle

\begin{abstract}

While the first kilometer-scale neutrino telescope, IceCube, is under construction, alternative plans exist to build even larger detectors that will, however, be limited by a much higher neutrino energy threshold of 10\,PeV or higher rather than 10 to 100 GeV. These future projects detect radio and acoustic pulses as well as air showers initiated by ultra-high energy neutrinos. As an alternative, we here propose an expansion of IceCube, using the same strings, placed on a grid with a spacing of order 500\,m. Unlike other proposals, the expanded detector uses methods that are understood and calibrated on atmospheric neutrinos. Atmospheric neutrinos represent the only background at the energies under consideration and is totally negligible. Also, the cost of such a detector is understood. We conclude that supplementing the 81 IceCube strings with a modest number of additional strings spaced at large distances can almost double the effective volume of the detector. Doubling the number of strings on a 800\,m grid can deliver a detector that this a factor of 5 larger for horizontal muons at modest cost.

\end{abstract}

%\keywords{gamma rays: bursts---acceleration of particles---neutrinos(fix these%)}

%%%%%%%%%%%%%%%%%%%%%%%%%%%%%%%%%%%%%%%%%%%%%%%%%%%%%%%%%%%%%%%%
\section{Introduction}

The first kilometer-scale neutrino observatory, IceCube, is now under construction\cite{icecube}. The instrumentation is based on the proven technology of the first-generation AMANDA-II telescope. It has taken data for more than 3 years and accumulated a $\sim 0.1\,km^2$\,year fluency of neutrinos with energy in the 50 GeV to 100 TeV range. IceCube is optimized to detect all flavors of neutrinos in the 100 GeV to 1 EeV energy range and identify their flavor over a large fraction of this energy range. Efforts exist \cite{review} to construct detectors with thresholds in the 10 PeV region and above to search for neutrino signals at the very highest energies such as those predicted to originate in Z-bursts \cite{zbursts}, topological defects related to GUT phase transitions in the early universe \cite{defects} and, generally, in top-down models for the origin of the highest energy cosmic rays. Such experiments employ a variety of novel techniques such as the detection of neutrino-induced horizontal air showers or the detection of the radio emission in the Giga-Hertz range by excess electrons in showers initiated by cosmic neutrinos \cite{radio}. Also the detection of the acoustic shock produced by the large amount of heat deposited by neutrino-induced showers has been investigated \cite{acoustic}.

Although optimized for TeV-PeV neutrinos, IceCube is also capable of observing neutrinos up to the scale of the highest energy cosmic rays \cite{alvarez}. At such high energies, showers and muons trigger photo-multipliers (PMTs) over hundreds of meters, much larger distances than the 125\,m string spacing planned for IceCube. Strings spaced over greater distances would be a far more economical method for studying ultra-high energy cosmic neutrinos.

It is the purpose of this paper to demonstrate how extensions of IceCube may be competitive with detectors specialized for the identification of neutrinos of 10 PeV energy and above. We propose expanding the IceCube detector beyond its 81 strings and cubic kilometer instrumented volume to a multi-kilometer detector optimized for EeV neutrino astronomy. We discuss two versions: ``IceCube-plus'' which surrounds IceCube with a ring of 13 to 18 conventional IceCube strings and ``Hypercube'' which roughly doubles the number of the strings in the current IceCube design. The additional strings will be spaced by distances between 300 and 1000 m, compared with the 125 m spacings of the original detector. Our proposal is competitive to other technologies in the sense that the large string spacings match the attenuation lengths of the acoustic and radio emission in ice.

Radio and acoustic detection are not necessarily alternatives to our proposal. Once the investment of
drilling the holes has been made, radio antennas and acoustic detectors can be deployed along with
IceCube digital optical modules (for recent discussions, see Ref.~\cite{radio,codeploy}). Each device
could assist with shower vertex location for both neutral and charged current events of all neutrino flavors. Vertex location is critical for energy
determination. Coincident observations would also allow an in-situ calibration of these techniques
with IceCube. This would resolve one of the often raised criticisms of radio and acoustic techniques.

In this context, the cosmogenic neutrino flux, sometimes called the GZK neutrino flux, is of special interest. These neutrinos are produced in the interactions between the ultra-high energy, extra-galactic, component of the cosmic ray spectrum and background microwave photons. These interactions are responsible for the Greisen-Zatsepin-Kuzmin (GZK) cutoff in the cosmic ray spectrum in the vicinity of 50 EeV. The number of GZK neutrinos detected are routinely used to characterize the performance of ultra-high energy neutrino experiments by a single number. While their observation may, arguably, be of moderate interest, they form an essentially guaranteed flux of neutrinos at the highest energies, with relatively minor uncertainties associated with the energy and redshift distribution of the sources \cite{cosmogenicunc}. Such a flux could be used to calibrate an experiment in the same way that atmospheric neutrinos are exploited to verify the performance of detectors such as AMANDA and IceCube. The characteristic energies of this flux fall in the range of 0.1 to 1\,EeV.

We find that supplementing the 81 IceCube strings with a modest number of additional strings, spaced at large distances, can almost double the effective volume of the detector. Doubling the number of strings on a 500-800\,m grid can deliver a detector that is a factor of 5 larger for horizontal muons.

%%%%%%%%%%%%%%%%%%%%%%%%%%%%%%%%%%%%%%%%%%%%%%%%%%%%%%%%%%%%%%%%

\section{Simulation}
\label{icecube}

IceCube\cite{PDD} will consist of 80 kilometer-length strings, each instrumented with 60 10-inch photo-multipliers spaced by 17~m.
The deepest module is 2.4~km below the surface. The strings are arranged at the apexes of equilateral triangles 125\,m on a side. The instrumented (not effective!) detector volume is a cubic kilometer that contains the AMANDA detector. To assess the ability of the IceCube detector and  its possible extensions to observe very high-energy neutrinos, we have designed a simple simulation to evaluate the response of a detector to showers (cascades) initiated by neutrinos of all flavors and to muon tracks initiated by $\nu_\mu$'s only. The latter can, however, reach a detector from distances of tens of kilometers at the energies of interest here.

\subsection{Showers}

\begin{figure}[t]
\begin{center}
\includegraphics[width=9.0cm,angle=90]{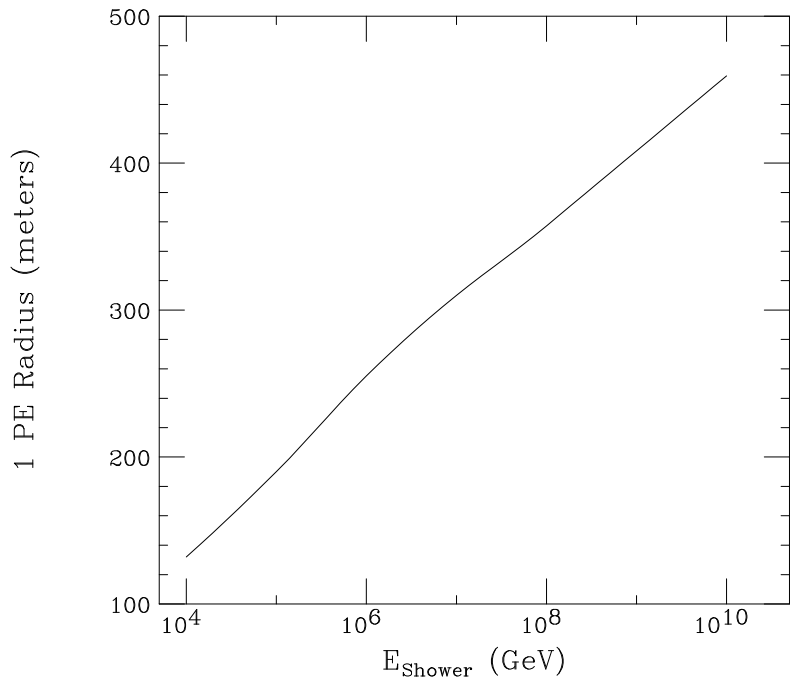}
\end{center}
\vspace*{0mm}
\caption[]{\label{fig:one}

The distance from a shower at which, on average, a single photo-electron is produced by a PMT embedded in deep ice as a function of shower energy. Its value roughly corresponds to the ideal string spacing for detecting showers of a given energy.}
\end{figure}

When a neutrino initiates an electromagnetic or hadronic shower in ice, Cherenkov photons are generated that can trigger PMTs over a volume whose radius increases with the shower energy. Although the Cerenkov light from electromagnetic and hadronic showers is preferentially emitted in the direction of the leading particles, for our considerations it can be well approximated as isotropic emission. Figure~\ref{fig:one} shows the radius of a shower produced as a function of energy for deep ice \cite{PDD,karle}. Throughout this article, ``shower'' refers to the leading shower produced by $\nu_e$ or $\nu_{\tau}$ in charged current interactions, as
well as the neutral current emission by all three flavors.

To be considered a detectable shower event, we require one of the following:
\begin{itemize}

        \item{} Two or more of the new strings have at least 170 meters of their length within the shower volume. This implies that at least 10 PMTs, on each of two strings, would report a signal given their 17\,m spacing.
 
        \item{} An (vertical) edge of the IceCube detector and at least one new string have at least 170 meters of their length within the shower volume.

	\item{} A vertical strip of the IceCube detector at least 120 meters inside of the edge has at least 170 meters of their length within the shower volume. We will call this class of event an ``IceCube-alone'' event. This simple simulation reproduces the expected performance of IceCube for ``IceCube-alone" events obtained with a complete detector simulation\cite{PDD}. 

\end{itemize}

The ideal detector geometry can now be selected by relating the shower size to the spacing of new strings. At the energies of interest here, PeV to EeV, typical shower sizes are on the order of several hundred meters. The symmetry of the shower event favors an even distribution of strings. This is not necessarily the case for muon events.

To determine the ideal spacing of any new strings, we have considered several geometries and calculated the new detector's effective shower volume as a function of shower energy. We define effective volume as:

\begin{equation}
V_{\rm{eff}}=\frac{N_{\rm{events}}}{N_{\rm{generated}}} \times V_{\rm{generated}},
\end{equation}
where $V_{\rm{generated}}$ is the volume over which showers are generated, much larger than the effective or geometric volumes. The effective volume is the equivalent geometric volume over which the detector is 100\% efficient for detecting interactions which occur inside its geometry.
      
\begin{figure}[t]
\begin{center}
\includegraphics[width=13.0cm,angle=90]{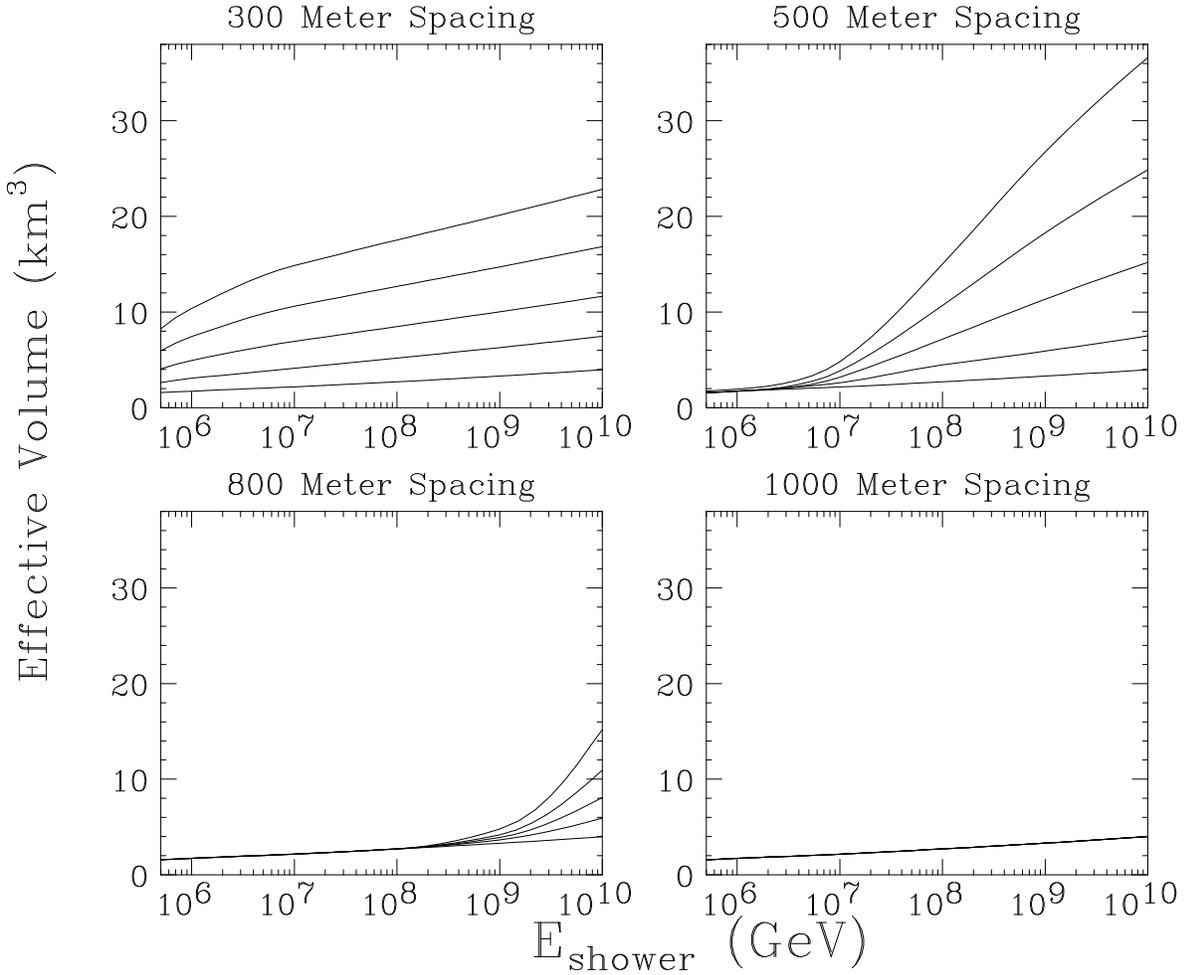}
\end{center}
\vspace*{0mm}
\caption[]{\label{fig:two}
The effective shower volumes as a function of shower energy for a variety of detector geometries. In each frame, from top to bottom represents geometries of 4, 3, 2, 1 and no (IceCube alone) added rings of strings. Note that for a string spacing of 1000 meters, the shower threshold energy for additional strings is greater than 10 EeV, and, therefore, the effective volumes for configurations with additional strings is indistinguishable to the effective shower volume for IceCube alone.}
\end{figure}

We have considered a variety of sample geometries: one to four rings of strings, with the strings within each ring separated by 300, 500, 800 or 1000 meters. The effective volumes for these configurations are shown as a function of energy in figure~\ref{fig:two}. In each of the four panels, the lowest line shown is for IceCube alone, with no added strings. The other four lines in each panel represent, from top to bottom, 4, 3, 2 and 1 additional rings of strings. We find that for showers, relatively modest spacings in the range of 300-500 meters are most favorable. Beyond this spacing, the energy threshold becomes prohibitively  high, even for detecting GZK neutrinos.

For 500 meter spacing, the addition of 91 strings (four rings) can improve the effective shower volume of IceCube by almost one order of magnitude at EeV energies. Alternatively, with only additional 13 strings, again separated by 500 meters, the effective volume is increased by a factor of 2.

\begin{figure}[t]
\begin{center}
\includegraphics[width=13.0cm,angle=90]{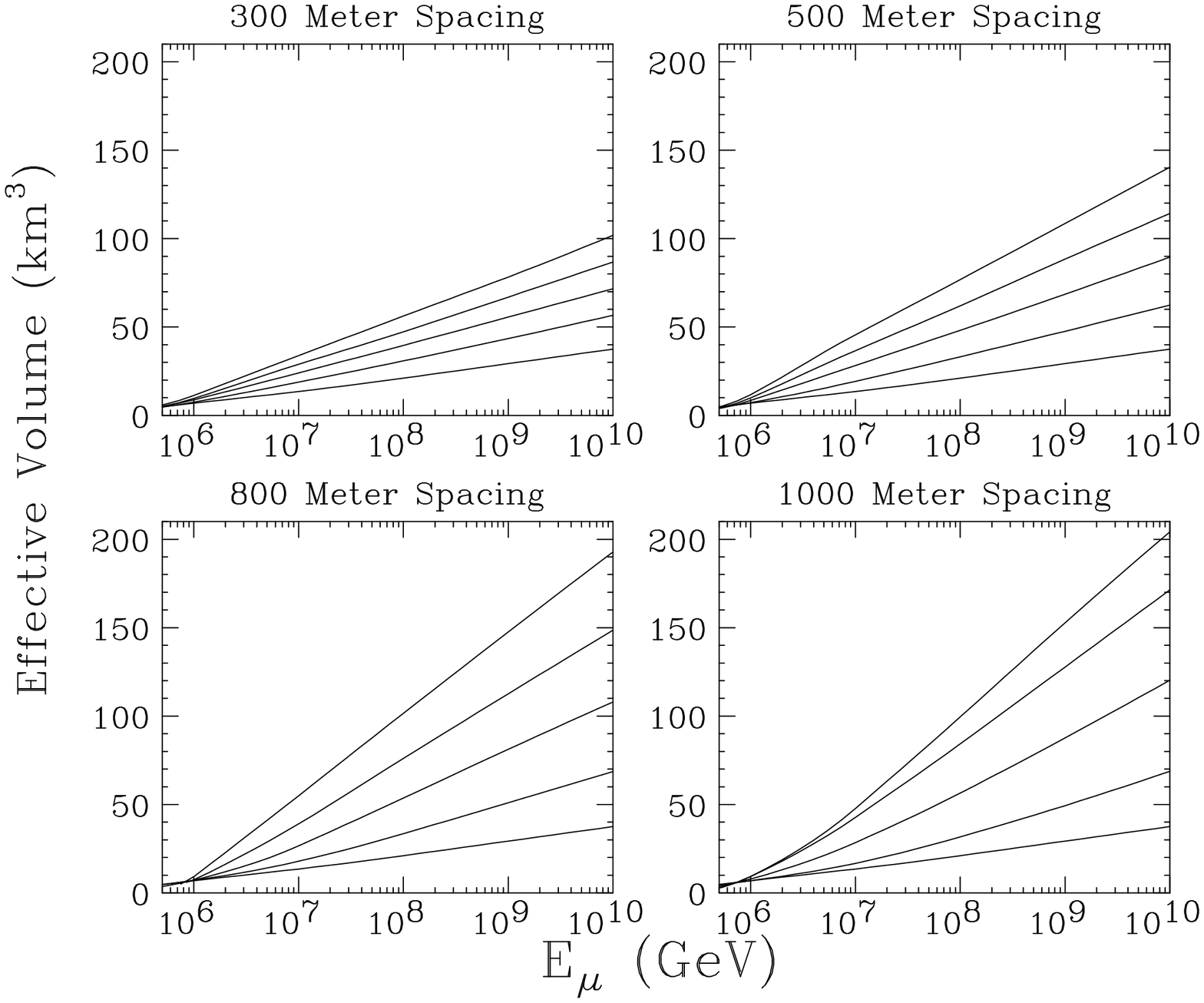}
\end{center}
\vspace*{0mm}
\caption[]{\label{fig:three}
 The effective volumes for horizontal muon events as a function of muon track energy for a variety of detector geometries. In each frame, from top to bottom represents geometries of 4, 3, 2, 1 and no (IceCube alone) added rings of strings.}
\end{figure}

\subsection{Muons}                

The simulation of muon events is somewhat more complicated. Muons travel many kilometers generating showers and Cherenkov light along their track by bremsstrahlung, pair production and photonuclear interactions\cite{prop,gaisser}. As the energy of the muon degrades, the distance from the track over which the light can trigger PMTs becomes smaller. The geometry of the volume surrounding the muon track over which single photo-electron are produced is a cone, many kilometers long.

High energy muons lose energy catastrophically according to\cite{prop}
\begin{equation}
\frac{dE}{dX}=-\alpha - \beta E \ ,
\end{equation}
where $\alpha=2.0 \times 10^{-6}~{\rm TeV}\, {\rm cm}^2/{\rm g}$ and $\beta=4.2 \times
10^{-6}~{\rm cm}^2/{\rm g}$ \cite{prop,gaisser}. The distance a muon travels before its energy drops below some energy threshold, $E^{\rm thr}_\mu$, called the muon range is then given by
\begin{equation}  
R_\mu = \frac{1}{\beta} \ln \left[ 
\frac{\alpha + \beta E_\mu}{\alpha + \beta E^{\rm thr}_\mu} \right]
\label{murange}
.
\end{equation}
At PeV to EeV energies, muons have ranges of tens of kilometers, greatly enhancing their detectability. We use similar detection requirements as we did for showers to evaluate the detector's effective volume for muons. For muons, however, unlike showers, the initial energy of the event cannot always be measured. A muon could be produced at one energy, travel several kilometers, and be detected with much less energy. For this reason, to be able to seperate ultra-high energy muon events from lower energy backgrounds, we must impose more stringent energy requirements. First, instead of requiring 170 meters of a string's length within a shower volume, we require 380 meters for muons. This corresponds to a $10^5\,$GeV shower, or a $\sim\,$PeV muon. Secondly, we require muons to have at least 1 PeV energy to trigger the original IceCube volume. Together, these requirements insure that all observed events are background free. 

In addition to selecting random locations for an event to occur, random incoming angles must be considered for muon simulation. Because of their absorption by the earth, it is sufficient to consider horizontal and downgoing muon events only. 

We first consider the same detector geometries as discussed for shower events. In figure \ref{fig:three} we show the effective volumes for these configurations.  These volumes are considerably larger than for showers because of the long muon ranges at the energies considered here. For all cases, spacings of 800-1000 meters yield the largest effective volumes for the detection of muons. With only 11 additional strings in a single ring of 800 meter radius, the effective volume is increased by a factor of almost 2 at EeV energies. Using four rings of strings (81 strings), the enhancement is considerably larger, about a factor of 5.

Given the symmetry of shower events, evenly distributed strings yield the greatest effective volume. This is not readily apparent for muons. For illustration, we have considered alternate geometries. First, consider a single ring of 80 new strings.  If we allow the radius of this ring to vary, we find that the effective muon volume becomes greatest (at 100 PeV) for a ring about 3.4 kilometers beyond IceCube.  This configuration has an effective muon volume of almost 120 $\rm{km}^3$, about ten percent larger than for the 800 meter, evenly spaced, distribution with the same number of additional strings. If two new rings, with a total of 80 additional strings are introduced, the maximum muon effective volume (at 100 PeV) is about 125 $\rm{km}^3$, for rings of 32 and 58 strings positioned 2.0 and 4.0 kilometers outside of IceCube, respectively. So, it is true that uneven distribution of strings can provide slightly larger effective volumes for muons. Such configurations are ineffective for showers, however.
 
\section{An Example: The Cosmogenic Neutrino Flux}

To illustrate the sensitivity of extended versions of IceCube, we will consider the detection of GZK neutrinos\cite{cosmogenic1}. This neutrino flux peaks near EeV energies, and therefore represents an interesting benchmark. The existence of this flux, called the cosmogenic neutrino flux, is a robust prediction although it does have a number of quantitative uncertainties associated with cosmological source evolution. We have used the flux described in Ref.\cite{cosmogenic2} as a somewhat conservative, but representative sample. 

Table I shows the event rates predicted from cosmogenic neutrinos for four choices of the detector configuration. The configuration, ``HyperCube'', refers to a design of four rings of strings, a total of 81 or 91 additional strings, separated by 800 or 500 meters, respectively. ``IceCube-Plus'' describes a more modest design of 13 or 18 new strings, separated by 500 or 300 meters, respectively. ``IceCube'', shown for comparison, refers to the original IceCube design, with no new strings added. Note that our rate for this configuration is in good agreement with the calculation of Ref.~\cite{alvarez}. ``$1\,\rm{km}^3$ Trigger Volume'' considers only showers which are created within the cubic kilometer instrumented volume of IceCube or showers which enter (above threshold) the instrumented volume, discounting large shower volumes which may trigger the detector from outside this volume.  

Notice that the small contribution of 13 to 18 new strings (IceCube-Plus) can enhance the associated event rate by a factor of almost 2.  The larger configuration (HyperCube) improves the rate by a factor of 4 or 5. The enhancement will be similar for other neutrino sources which peak at EeV scales, including neutrinos from top-down cosmic ray scenarios \cite{topdown}, for example.

For comparison, we consider examples of future experiments exclusively sensitive to ultra-high energy neutrinos. First, the ANITA experiment \cite{anita}, a balloon-borne radio antenna array designed to study ultra-high energy neutrinos interacting in Antarctic ice. Taking advantage of its high altitude position, ANITA samples an enormous effective volume of $10^6$ cubic kilometers of ice at EeV energies. Only earth-skimming events are observed, however, which limits ANITA's acceptance to $\sim 10^{-2}$ steradians.  Over a ten day flight, ANITA is expected to observe about 1 cosmogenic neutrino \cite{anita}.

The proposed SALSA experiment uses large salt domes as a radio Cherenkov medium \cite{salsa}. With $10^4$ times less target mass than ANITA, SALSA can compensate for this with a full $2\pi$ steradians acceptance and continuous operation. In one year of operation, SALSA will observe the same number of cosmogenic neutrino events as ANITA in 25 days of flight time. We have here assumed a SALSA design with $8 \, \rm{km}^3$ effective volume and $2.2 \, \rm{g}/\rm{cm}^3$ of target density. 

Prior to SALSA or ANITA, there will be a number of other experiments potentially capable of observing EeV neutrinos, including AUGER \cite{auger}, ANTARES \cite{antares} and RICE \cite{rice}. Despite the advances such experiments represent, efforts beyond these will be required to more throughly explore the EeV neutrino spectrum.

Extensions of IceCube will also impact the experiment's ability to do neutrino astronomy at more modest energies, perhaps at the PeV scale. Although the extensions proposed here are not optimized for PeV energies, enhancements of a factor of 2 to 4 are still possible. Many interesting PeV neutrino sources are likely to exist. These include, but are not limited to, gamma-ray bursts \cite{grb} and active galactic nuclei \cite{agn}. Additionally, unlike other experiments, IceCube and its extensions have the ability to identify the flavor of high-energy cosmic neutrinos observed \cite{flavor}. 

Prospects for the detection of tau neutrinos, via their ``double-bang'' signature \cite{learnedpakvasa}, could also be enhanced with an extension of IceCube. In such an event, a tau neutrino interacts within the detector producing a tau lepton and a shower. The tau lepton then travels away from the shower where it decays, producing a second shower, ideally also within the detector volume. At PeV energies, the separation of these showers is on the order of 500 meters, and the experiments proposed in this paper could potentially observe such events. At GZK energies, however, the two showers will be separated by several kilometers and such observations will be difficult.

%%%%%%%%%%%%%%%%%%%%%%%%%%%%%%%%%%%%%%%%%%%%%%%%%%%%%%%
\vspace{0.0cm}
 \begin{table}
 \label{table1}
 
 \hspace{1.5cm}
 \begin{tabular} {c c c c} 
 & Showers & Muons & \\
 \hline \hline
%$1\,\rm{km}^3$ Trigger Volume  & 0.19 & 0.91 & \\
IceCube & 0.60 & 0.76 & \\
%\hline
IceCube-Plus (300 m) & 1.1 & 1.1 & \\
IceCube-Plus (500 m) & 1.1 & 1.2 & \\
%\hline
HyperCube (500 m) & 4.5 & 2.8 & \\
HyperCube (800 m) & 1.2 & 3.8 & \\
% \hline \hline
 \end{tabular}
 \caption{The event rate (per year) for a variety of detector configurations, for both shower and muon events. A representative cosmogenic neutrino flux has been used (see text). Note that the modest IceCube extension, ``IceCube-Plus'' nearly doubles the event rate predicted for IceCube and the larger, ``HyperCube'' configuration enhances the event rate by a factor of about 5.}
 \end{table}
%%%%%%%%%%%%%%%%%%%%%%%%%%%%%%%%%%%%%%%%%%%%%%%%%%%%%%%%%%

%%%%%%%%%%%%%%%%%%%%%%%%%%%%%%%%%%%%%%%%%%%%%%%%%%%%%%%
\vspace{1.0cm}
 \begin{table}
 \hspace{1.5cm}
 \begin{tabular} {c c c} 
 & Events & \\
 \hline \hline
ANITA-10 Days (1 Flight)  & $\sim\,$1 & \\
ANITA-30 Days (3 Flights) & $\sim\,$3 & \\
SALSA                     & $\sim\,$3 (per year) & \\
% \hline \hline
 \end{tabular}
 \caption{The event rates for other proposed/planned ultra-high energy neutrino experiments. The same cosmogenic neutrino flux has been used as for the IceCube-Plus and HyperCube calculations (see text). Note that IceCube-Plus has similar rates compared to these experiments and HyperCube has considerably higher rates.} 
 \end{table}
%%%%%%%%%%%%%%%%%%%%%%%%%%%%%%%%%%%%%%%%%%%%%%%%%%%%%%%%%%

\section{Conclusions}

With the addition of new strings, sparsely spaced around the planned IceCube experiment, a very large volume, ultra-high energy neutrino observatory could be developed. With less than 20 new strings, IceCube's sensitivity to EeV scale neutrinos can almost be doubled. With 80 new strings, an enhancement of a factor of 5 can be accomplished .

Comparing these configurations to other future ultra-high energy neutrino experiments, ANITA and SALSA, we find that an extension of IceCube would be a competitive observatory. Given the existence of the IceCube detector and its associated infrastructure, it should be clear that these extensions could be achieved at modest cost.

\acknowledgements

We would like to thank Albrecht Karle, David Besson, Peter Gorham, John Learned and David Saltzberg for valuable discussions. 
This research was supported  by the U.S.~Department of Energy
under grant DE-FG02-95ER40896
and by the Wisconsin Alumni Research Foundation.

\end{document}

%% file: declare.tex
% macros for marking changes
\marginparwidth 1.cm
\setlength{\hoffset}{-1cm}
\newcommand{\mpar}[1]{{\marginpar{\hbadness10000%
                      \sloppy\hfuzz10pt\boldmath\bf\footnotesize#1}}%
                      \typeout{marginpar: #1}\ignorespaces}
\def\mda{\mpar{\hfil$\downarrow$\hfil}\ignorespaces}
\def\mua{\mpar{\hfil$\uparrow$\hfil}\ignorespaces}
\def\mla{\marginpar[\boldmath\hfil$\rightarrow$\hfil]%
                   {\boldmath\hfil$\leftarrow $\hfil}%
                    \typeout{marginpar: $\leftrightarrow$}\ignorespaces}

\def\ba{\begin{eqnarray}}
\def\ea{\end{eqnarray}}
\def\bq{\begin{equation}}
\def\eq{\end{equation}}

\renewcommand{\abstractname}{Abstract}
\renewcommand{\figurename}{Figure}
\renewcommand{\refname}{Bibliography}

% peter's conventions 
\newcommand{\eg}{{\it e.g.}\;}
\newcommand{\ie}{{\it i.e.}\;}
\newcommand{\etal}{{\it et al.}\;}
\newcommand{\ibid}{{\it ibid.}\;}

% additional commands 
\newcommand{\mx}{M_{\rm SUSY}}
\newcommand{\pt}{p_{\rm T}}
\newcommand{\et}{E_{\rm T}}
\newcommand{\del}{\varepsilon}
\newcommand{\sla}[1]{/\!\!\!#1}
\newcommand{\fb}{\;{\rm fb}}
\newcommand{\pb}{\;{\rm pb}}
\newcommand{\mev}{\;{\rm MeV}}
\newcommand{\gev}{\;{\rm GeV}}
\newcommand{\tev}{\;{\rm TeV}}
\newcommand{\abi}{\;{\rm ab}^{-1}}
\newcommand{\fbi}{\;{\rm fb}^{-1}}

\newcommand{\lsusy}{\lambda'_{\rm SUSY}}
\newcommand{\llq}{\lambda'_{\rm LQ}}
\newcommand{\msbar}{\overline{\rm MS}}
\newcommand{\met}{E\hspace{-0.45em}|\hspace{0.1em}}

% all the susy stuff
\newcommand{\SP}{\scriptscriptstyle}
\newcommand{\stl}{\tilde{t}_{\SP L}}
\newcommand{\str}{\tilde{t}_{\SP R}}
\newcommand{\ste}{\tilde{t}_1}
\newcommand{\stz}{\tilde{t}_2}
\newcommand{\st}{\tilde{t}}
\newcommand{\gt}{\tilde{g}}
\newcommand{\sle}{\tilde{\tau}_1}
\newcommand{\slz}{\tilde{\tau}_2}
\newcommand{\che}{\tilde{\chi}^\pm_1}
\newcommand{\cpe}{\tilde{\chi}^+_1}
\newcommand{\cme}{\tilde{\chi}^-_1}
\newcommand{\nne}{\tilde{\chi}^0_1}
\newcommand{\nnz}{\tilde{\chi}^0_2}
\newcommand{\mse}{m_{\tilde{t}_{\SP 1}}}
\newcommand{\msz}{m_{\tilde{t}_{\SP 2}}}
\newcommand{\mst}{m_{\tilde{t}}}
\newcommand{\mg}{m_{\tilde{g}}}
\newcommand{\ms}{m_{\tilde{q}}}
\newcommand{\mt}{m_t}
\newcommand{\mheavy}{m_{\rm heavy}}
\newcommand{\mle}{m_{\tilde{\tau}_{\SP 1}}}
\newcommand{\mce}{m_{\tilde{\chi}^+_{\SP 1}}}
\newcommand{\mne}{m_{\tilde{\chi}^0_{\SP 1}}}